# GEM operation in helium and neon at low temperatures

A. Buzulutskov [a*], J. Dodd [b], R. Galea [b], Y. Ju [b],
M. Leltchouk [b], P. Rehak [c], V. Tcherniatine [c], W. J. Willis [b],
A. Bondar [a], D. Pavlyuchenko [a], R. Snopkov [a], Y. Tikhonov [a]

[a] *Budker Institute of Nuclear Physics, Novosibirsk 630090, Russia*
[b] *Nevis Laboratories, Columbia University, Irvington, NY 10533, USA*
[c] *Brookhaven National Laboratory, Upton, NY 11973, USA*

**Abstract**

We study the performance of Gas Electron Multipliers (GEMs) in gaseous He, Ne and Ne+$H_2$ at temperatures in the range of 2.6-293 K. In He, at temperatures between 62 and 293 K, the triple-GEM structures often operate at rather high gains, exceeding 1000. There is an indication that this high gain is achieved by Penning effect in the gas impurities released by outgassing. At lower temperatures the gain-voltage characteristics are significantly modified probably due to the freeze-out of impurities. In particular, the double-GEM and single-GEM structures can operate down to 2.6 K at gains reaching only several tens at a gas density of about 0.5 g/l; at higher densities the maximum gain drops further. In Ne, the maximum gain also drops at cryogenic temperatures. The gain drop in Ne at low temperatures can be reestablished in Penning mixtures of Ne+$H_2$: very high gains, exceeding $10^4$, have been obtained in these mixtures at 50-60 K, at a density of 9.2 g/l corresponding to that of saturated Ne vapor near 27 K. The results obtained are relevant in the fields of two-phase He and Ne detectors for solar neutrino detection and electron avalanching at low temperatures.

*Keywords:* Gas electron multipliers; Cryogenic avalanche detectors; Helium; Neon; Penning mixtures.
*PACs:* 29.40.Cs; 34.80.My.

## 1. Introduction

The studies of Gas Electron Multipliers (GEMs) [1] at cryogenic temperatures have been motivated by a growing interest in the development of cryogenic detectors operated in an electron-avalanching mode. Such detectors, termed "cryogenic avalanche detectors" [2], may find applications in cryogenic experiments where the primary ionization signal is weak: in solar neutrino detection [3], dark matter searches [4], coherent neutrino scattering [5] and Positron Emission Tomography [6].

The unique property of GEM structures is that they can operate in noble gases at high gains [7,8], including at cryogenic temperatures. Indeed, it has been recently demonstrated that the triple-GEM [2,9] and single-GEM [10] structures can successfully operate in gaseous He, Ar and Kr and in two-phase Kr in the range of 120-300 K.

---

[*] Corresponding author. Email: buzulu@inp.nsk.su

The interest in GEM performance at low temperatures, below the liquid nitrogen point (78 K), has been triggered by the so-called "e-bubble chamber" project for solar neutrino detection [3,11]. In this project, the proposed two-phase detector should operate in those liquids that could provide the formation of electron bubbles [12], namely in neon, hydrogen or helium. The corresponding operation temperatures would be below 27, 20 or 4.2 K. Furthermore, the detector should be able to amplify the ionization signal produced in the liquid using a gas multiplier, for instance the GEM, operated in a saturated vapor above the liquid phase. The latter is a real challenge. For example, the saturated He vapor density at 4.2 K (16 g/l) corresponds to the gas density at room temperature at a pressure of 101 atm. Very little is known about electron avalanching at low temperatures and at such a high density [13].

To the best of our knowledge, only three groups have experimental results relevant to this subject: those of the GEM performance in He at cryogenic temperatures [2,9,10], the proportional counter operation in He near 4 K [14,15] and the cryogenic glow discharge in He [16].

The first group observed that the triple-GEM can operate in He at high gains (exceeding $10^4$) down to 120 K and that the gain-voltage characteristics are independent of the temperature [9]. Based on the pressure dependence of ionization coefficients, it was supposed that associative ionization (the Hornbeck-Molnar process [17]) is the mechanism responsible for high gain operation of GEMs in dense He (and Ne), at gas densities exceeding 0.5 g/l [18].

The second group states that associative ionization should be suppressed at low temperatures and that electron impact ionization remains the only avalanche mechanism near 4 K [15]. The maximum gain, obtained in a single-wire proportional counter near 4 K at a He density of 0.2 g/l, was 200.

The third group pointed to the crucial role of metastable He atoms created in the discharge. It was supposed that collisions between them, followed by ionization, results in a change of the discharge mechanism at low temperatures [16]. All groups emphasize the importance of He ion clustering at low temperatures, an effect that might also modify the avalanche mechanism.

In this paper, the performance of GEM structures in gaseous He, Ne and Ne+$H_2$ are studied for the first time at low temperatures, down to 2.6 K, aiming at their potential applications in the e-bubble chamber and at investigations of electron avalanching at low temperatures. Gain characteristics, pulse shapes and interpretation of the results are presented.

## 2. Experimental procedure

A detailed description of the experimental setup will be presented elsewhere [19]. The setup was developed at Nevis Laboratories (Columbia University) and Brookhaven National Laboratory. Fig. 1 shows the principal features of the setup relevant to the present study. Three GEM foils and a photocathode were mounted in a test vessel of a volume of 1.6 l, termed the "tracking chamber", which was placed inside the cryostat. The chamber was connected to a tube which was in direct contact with a reservoir filled with a cryogenic liquid. In the range of 78-290 K the reservoir was filled with liquid nitrogen, and below 78 K with liquid helium. The chamber was filled with appropriate gases, He, Ne or $H_2$, taken directly from the bottles, with quoted purities of 99.999%.

The photocathode was composed of a semitransparent gold film deposited on a quartz substrate. The chamber and the cryostat had UV windows at the bottom, to transmit the light to the photocathode from a high-powered Xe flash lamp, placed underneath. The lamp was pulsed once per second using a generator, which also served as a trigger for the signal readout. The width of the light pulse was approximately 2 μs.

The GEMs were produced by the CERN workshop and had the following parameters: 50 μm thick kapton, 70 and 55 μm hole diameter on the metal and kapton center respectively, 140 μm hole pitch and 28×28 mm$^2$ active area. The GEM foils were mounted on G10 frames. The distances between the first GEM (GEM1 in Fig. 1) and the photocathode, and between the GEMs, were 10 mm and 2 mm, respectively.

The GEM electrodes were biased through a resistive high-voltage divider placed outside the cryostat and including three identical circuits connected in parallel (Fig.2). The divider was slightly modified compared to that of Ref. [9], to provide better protection against discharges. Each GEM electrode was connected to the divider using high-voltage feedthroughs and 3 m long wires. All resistors in the divider were equal, providing an equal voltage drop across the GEMs and the gaps between them.

The GEM multiplier could operate in a single-, double- or triple-GEM configuration, denoted as 1GEM, 2GEM or 3GEM respectively, in accordance with terminology of Ref. [7]. Accordingly, the avalanche (anode) signal was recorded from the last electrode of the first, second or third GEM. Correspondingly, the preamplifier and the capacitors C1 and C2 (their function is explained in Ref. [10]) shown in Fig.2 were connected to the appropriate GEM wires. In the double-GEM, the third GEM was disconnected from the divider. In the single-GEM, both the third and second GEMs were disconnected.

The "calibration" signal was induced by electrons arriving at the first GEM from the photocathode, i.e. before avalanche amplification. It was recorded at the first electrode of the first GEM, with no high voltage applied on the divider.

Both avalanche and calibration signals were read out using a charge-sensitive preamplifier followed by a research amplifier, the latter providing a varying shaping time for the pulse, in the range of 0.5-10 μs. The gain value is defined as the pulse-height of the avalanche signal divided by that of the calibration signal, at a shaping time of 10 μs.

The maximum gain is defined as that limited by the onset of either a sustained discharge or discharge-like signals. For most of the data presented in the following figures, the maximum gain was not reached, unless otherwise specified.

In the present study, the drift field in the photocathode gap was kept equal to 1 kV/cm at all times.

### 3. GEM operation in He

Fig.3 shows gain-voltage characteristics of the triple-GEM in He in the range of 39-293 K. Most data were obtained at a gas density of about 0.5 g/l, corresponding to a pressure of 3 atm at room temperature. The data at 293 K were obtained at room temperature when the cryostat reservoir was empty. The data in the range of 78-240 and below 40 K were obtained during the cooling-down procedure, when the reservoir was filled with liquid nitrogen and liquid helium respectively. The data at 62 K were obtained during the warming-up procedure, after liquid helium was removed from the reservoir.

One can see that in the range of 62-238 K the characteristics are practically identical. At first glance this confirms the results obtained in Refs. [2,9], namely the temperature independence of the gain in He. In contrast, the gain-voltage characteristics move towards higher voltages when going from 293 to 238 K and from 62 to 39 K. Note that the maximum gains in Fig.3 (and Fig.4) are shown for the data obtained at 39 and 4.2 K only. For other temperatures the maximum gains were not reached: they were far higher than 1000.

The transition from 62 K to lower temperatures is characterized by a significant increase of the voltage and by a dramatic decrease of the maximum gain. At 39 K the maximum gain dropped considerably, below 10. Here the maximum gain was limited by the onset of a sustained discharge. We found that due to this fact the triple-GEM was unable to operate at temperatures below 39 K. Presumably, ion feedback between GEM elements is responsible for this behavior in the triple-GEM at low temperatures.

In contrast, the double-GEM and single-GEM were able to operate at low temperatures in an avalanche mode (see Figs.3-5). Fig.4 shows gain-voltage characteristics of the single-GEM and double-GEM at 4.2 K in comparison with those at 78 K. At lower temperatures the operation voltages increase and the curve slopes decrease, indicating a change in the avalanche mechanism when going from liquid nitrogen to liquid helium temperatures. On the other hand, no difference in gain characteristics was observed below 20 K. This is seen in particular in Fig. 5 when comparing the data at 4.4 and 2.6 K.

A typical avalanche signal from the double-GEM in He at 4.2 K is shown in Fig. 6. One can see that the pulse width of the avalanche (primary) signal is practically identical to that of the calibration signal. That means that the avalanche signal in He at low temperatures is faster than the shaping time of the amplifier (1.5 µs). The avalanche signal is slightly delayed with respect to the calibration signal, by 1.3 µs. This presumably corresponds to the electron drift time from the first to second GEM.

At higher gains the primary signal starts to be accompanied by a secondary signal, with a width of a few tens of microseconds at 4.2 K (Fig. 6, bottom). It is interesting that at lower temperatures, namely at 2.6 K, the width of the secondary signal increases by two orders of magnitude, up to a few milliseconds (Fig. 7, top, middle). The origin of the secondary signals is unclear. It is possibly related to long-lived metastable atoms created in an avalanche. In principle they might be responsible for ionization processes at such a large time-scale.

At 4.2 K and at a density of 0.55 g/l, the maximum gains of the single-GEM and double-GEM were approximately 20 and 60 respectively. Although at these gains we did not observe sustained discharges, an appreciable number of signals had a discharge-like shape, with a considerably larger width of the primary signal and larger contribution of the secondary signal. An example of the discharge signal is shown in Fig. 7 (bottom).

The density effect on gain-voltage characteristics at low temperatures is illustrated in Fig. 8. The maximum density reached was 1.6 g/l, at 2.6 K. This density corresponds to that of saturated He vapor at 2.3 K. One can see that the gain curve slope decreases with increasing density. In addition, the maximum gain at 1.6 g/l falls to a rather small value, of about 5 for the double-GEM. Such a strong density dependence of the maximum gain presumably imposes a limit on the low temperature performance of GEMs at high densities, above 1 g/l. This means that GEM structures can hardly operate in saturated vapor in the two-phase mode at 4.2 K, with a reasonable gain. On the contrary, low densities, below about 0.5 g/l, may provide good GEM performance at high gains. For example, the density value of 0.6 g/l, at which the double-GEM has a gain of several tens, corresponds to the density of saturated He vapor at 1.9 K. The saturated vapor density could also be reduced further, by an order of magnitude, if the temperature is lowered down to 1.3 K. At such a low density one might expect a significant increase of the GEM gain.

Charging-up effects induced just by low temperatures were not observed, in accordance with earlier observation [9], despite the fact that the kapton resistivity is extremely large at these temperatures. In particular, gain-voltage characteristics do not deviate from exponential behavior.

In addition, no fatigue effects were observed between the cooling-warming runs which were done during the present study. Some GEM elements were changed between the runs and some were left. The results nevertheless were reproducible. These observations demonstrate that the GEM structures can operate down to liquid helium temperatures.

## 4. GEM operation in Ne

Fig.9 (top, open points) shows a gain-voltage characteristic of the triple-GEM in Ne at room temperature, at a density of 4.2 g/l corresponding to a pressure of 5 atm. According to previous results [18], the GEM operation at this density is efficient in terms of the maximum gain. Indeed, the maximum gain in Fig. 9 exceeds 1000. However, below room temperature, namely at 160 and 78 K, we found that the operation voltages increased and that the triple-GEM was unable to operate in amplification mode due to the discharge onset. This behavior resembles that observed in He at temperatures below 40 K.

Fortunately, neon has an advantage over helium: it can form Penning mixtures with hydrogen at low temperatures. Indeed, the boiling point of hydrogen is below that of neon. It is known that Ne and $H_2$ constitute a Penning pair where the energy of the metastable state of neon ($Ne^m$) exceeds the ionization energy of hydrogen, resulting in a significant enhancement of the ionization coefficient [20] due to the following reactions:

$$e + Ne \rightarrow e + Ne^m$$
$$Ne^m + H_2 \rightarrow Ne + H_2^+ + e \qquad (1)$$

Thus, the problem of the gain drop at low temperatures might be solved using the Penning mixture Ne+$H_2$, which should be effective down to 20 K.

Fig. 9 shows gain-voltage characteristics of the single-GEM, double-GEM and triple-GEM at 55-57 K in the Penning mixtures Ne+0.2%$H_2$ and Ne+0.1%$H_2$. The gas densities were 4.1 g/l and 9.2 g/l respectively. Note that the latter density value is close to that required for operation in the two-phase mode near 27 K. The measurements were done during the warming-up procedure, after liquid helium was removed from the reservoir.

One can see that rather high gains, exceeding $10^4$, are obtained in these mixtures: no discharges were observed at these gains. We believe that such a successful performance of Ne+$H_2$ mixtures will be preserved at lower temperatures, down to 27 K.

High gain operation of the triple-GEM in Ne+0.1%$H_2$ is further illustrated in Fig. 10, showing a typical avalanche signal. One can see that the avalanche signal is slower than that in He at low temperatures (Fig. 6). It is wider, by approximately 2 μs, and delayed, by 5.1 μs, with respect to the calibration signal. The delay is determined by the electron drift time between the GEMs and by the avalanche development time within the GEM holes. The latter will be discussed in detail in the following section.

## 5. Discussion

We tried to understand the temperature dependence of the gain-voltage characteristics observed in He and Ne by analyzing their ionization coefficients. They were estimated using an approach described in Ref. [18]. Each GEM hole is approximated by a parallel-plate counter with an electric field taken equal to that calculated in the center of the hole [21]. The ionization coefficient $\alpha$ is related to the single-GEM gain $M$ as follows:

$$\alpha / N = \ln M / (Nd) \quad , \qquad (2)$$

where $N$ is the atomic density, $d$ the hole height. In the case of the triple-GEM with gain

$G$, the single-GEM gain is estimated using the formula

$$G = M^3 \varepsilon^2, \qquad (3)$$

where $\varepsilon$ is the charge transfer efficiency from GEM output to the following elements [21].

Fig. 11 shows a comparison of ionization coefficients in He at low temperatures, estimated using our data from Figs. 4 and 8, with those at room temperature taken from literature [22]. Note that the latter coefficients were obtained at rather low gas density. One can see that the data at temperatures below 20 K are in reasonable agreement with those at room temperature. This suggests that the avalanche mechanism at low temperatures and relatively high density (above 0.3 g/l) is similar to that at room temperature and low densities (below 0.02 g/l), i.e. it is described by electron impact ionization.

On the other hand, the ionization coefficients at temperatures above 60 K obtained from our data, including those at room temperature (not shown), are significantly enhanced (Fig. 11). Two questions arise: which mechanism (or mechanisms) is responsible for this enhancement and why it is suppressed at low temperatures.

In Ref. [18] the enhancement of ionization coefficients at room temperature and high pressures was interpreted as the result of the associative ionization:

$$e + He \rightarrow e + He^*$$
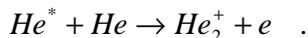
$$He^* + He \rightarrow He_2^+ + e \quad . \qquad (4)$$
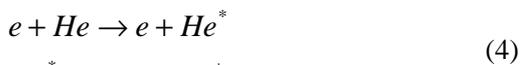

The interpretation was based on the observed pressure dependence of ionization coefficients. Namely, the contribution of associative ionization should increase with gas density in proportion to the density. At low temperatures, the associative ionization contribution should decrease, in proportion to or faster than $T^{1/2}$ [15].

We suppose that in the present study an alternative mechanism for this enhancement, namely the Penning ionization of impurities, might play a decisive role. Indeed, there is a correlation between the increase of the operation voltage and the cooling medium used in the cryostat reservoir (Fig. 3). The transition from 293 to 238 K is provided by liquid nitrogen and from 78 to 39 K by liquid helium filling. Accordingly, the tube in Fig.1 may work as a cold trap removing $CO_2$ and $H_2O$ impurities at liquid nitrogen filling and Ar, $O_2$ and $N_2$ impurities at liquid helium filling. The correlation might be explained by the presence of these impurities in He or Ne which might induce the Penning effect. Their removal would result in an increase in the operation voltage. In this sense, only the measurements at low temperatures, below 40 K, would be in really pure He.

Fig. 12 shows a comparison of ionization coefficients in Ne at room temperature and in Ne+$H_2$ at low temperatures estimated from our data, with those at room temperature and low density taken from literature [23]. Similar to the He case, the ionization coefficients in Ne deduced from our measurements are enhanced. The enhancement is higher in Penning mixtures Ne+(0.1-0.2)%$H_2$.

Long-lived metastable states play a crucial role in the Penning mechanism. Evidence for their presence in an avalanche was obtained by analyzing the time properties of the signals. Indeed, it was observed that the avalanche signals from the triple-GEM in He at 62 K and in Ne at room temperature are substantially delayed with respect to the calibration signal. The delay is of the order of 10 μs. This is seen from Fig. 13, which should be compared to calibration signal in Fig. 6. The delay is too large to be explained by the electron drift time between the GEMs, and thus can be induced only by the avalanche development time.

It is interesting that the delay is a logarithmic function of the gain, as is clearly seen in Fig. 14. A simple model can explain this dependence. Let $\tau$ be the life-time of an excited atom before ionization collision. The avalanche time in a single GEM is approximately equal to $\tau N_{GEN}$, where $N_{GEN}$ is the number of multiplication generations in the avalanche, equal to $\alpha d$. In the case of the triple-GEM, we obtain the avalanche time in all three GEMs as a logarithmic function of the gain, accounting for expressions (2) and (3):

$$T \approx 3\tau N_{GEN} = 3\tau\alpha d = 3\tau \ln M = \tau(\ln G - 2\ln\varepsilon) \quad . \tag{5}$$

Thus, the line slope in Fig. 14 provides an estimation of the life-time of an excited atom: $\tau = 2.4$ μs.

It is apparent that only metastable atoms can live such a long time. Concerning the Penning mechanism, estimations showed that the $N_2$ impurity concentration corresponding to this life-time would be about $5 \times 10^{-5}$. Another conclusion is that the contribution of associative ionization at temperatures near 78 K is ruled out (at least in the present studies), since the associative ionization is a fast process going through the resonance atomic states with a typical decay time below $10^{-7}$ s [24,25].

At room temperature, the signal in He is fast in contrast to temperatures near 78 K (see Fig.13). Regarding the Penning mechanism, one might suppose that the impurity concentration here was so high that the life-time of metastable atoms before ionization collisions was strongly reduced. The associative ionization contribution cannot be ruled out here either.

We also cannot rule out another Penning process which can occur in pure He (or in pure Ne):

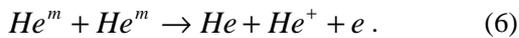
$$He^m + He^m \rightarrow He + He^+ + e \quad . \tag{6}$$

In Ref. [16] this process was considered as the principle mechanism of the glow discharge at temperatures near and below 78 K. Two facts support this mechanism. First, it goes through metastable atoms, which we know exist in the avalanche. Second, its contribution at low temperatures should decrease [25]. It is not clear however if the concentration of metastable atoms in the avalanche would be enough to start the process. This will be estimated elsewhere.

To conclude the discussion section, further measurements are needed in order to understand the avalanche mechanism in He and Ne near and above 78 K, preferably performed in cleaner conditions and with more stringent monitoring of gas purity.

## 6. Conclusions

We have studied the GEM performance in gaseous He, Ne and Ne+$H_2$ at low temperatures, in the range of 2.6-293 K. In the range of 62-293 K in He, the triple-GEM structures can operate at rather high gains, exceeding 1000. In the range of 62-238 K, the gain-voltage characteristics are independent of the temperature. At lower temperatures however they are significantly modified: the operation voltages increase and the maximum gains decrease. In particular, the double-GEM and single-GEM structures can operate down to 2.6 K at gains reaching only several tens, at a gas density of about 0.5 g/l. At higher densities the maximum gain drops even more. In Ne, the maximum gain also drops at cryogenic temperatures.

In Ne, the problem of the gain drop at low temperatures has been solved using Penning mixtures Ne+(0.1-0.2)%$H_2$. Very high gains, exceeding $10^4$, were obtained in these mixtures at 50-60 K, at densities corresponding to that of saturated Ne vapor at 27 K.

At temperatures below 20 K, the ionization coefficients in dense He are in accordance with those measured at room temperature and low densities. This supports the statement that the principal avalanche mechanism in He at low temperatures is electron impact ionization.

On the other hand, the avalanche mechanism in He and Ne near and above 78 K is not fully understood. Most probably it is due to Penning ionization of impurities, although another Penning process, due to collisions between metastable atoms, is not excluded.

There were no charging-up and fatigue effects observed. This indicates that the GEM structures can successfully operate down to liquid He temperatures.

The results obtained are relevant for understanding basic mechanisms of electron avalanching at low temperatures and for applications in two-phase He and Ne detectors for solar neutrino detection. Further studies of this technique are in progress.

## Acknowledgements

The research described in this publication was made possible in part by Award RP1-2550-NO-03 of the U.S. Civilian Research & Development Foundation for the Independent States of the Former Soviet Union (CRDF), and in part by Award PHY-0098826 of the U.S. National Science Foundation.

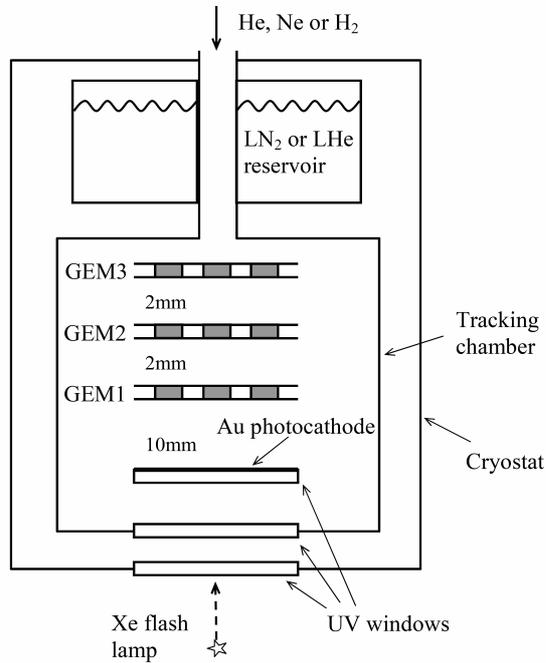

Fig.1 Schematic view of the experimental setup to study GEM performance at low temperatures (not to scale).

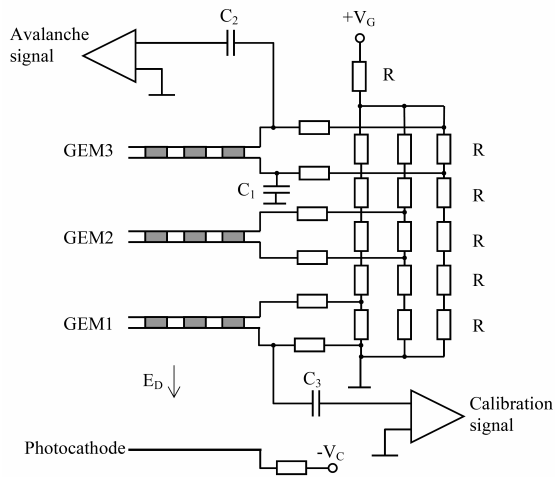

Fig.2 Electrical connections of the triple-GEM to a high-voltage divider and readout electronics. The value of each resistor is 10 MΩ.

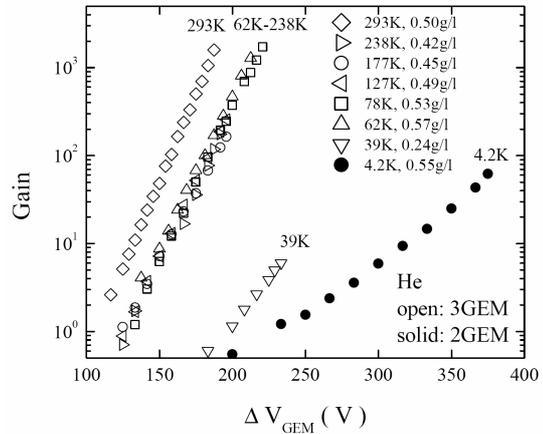

Fig.3 Gain-voltage characteristics of the triple-GEM in He in the range of 39-293 K and that of the double-GEM at 4.2 K. Gains as a function of the voltage across each GEM are shown. The appropriate temperatures and gas densities are indicated. At 4.2 and 39 K the maximum gains are limited by discharges.

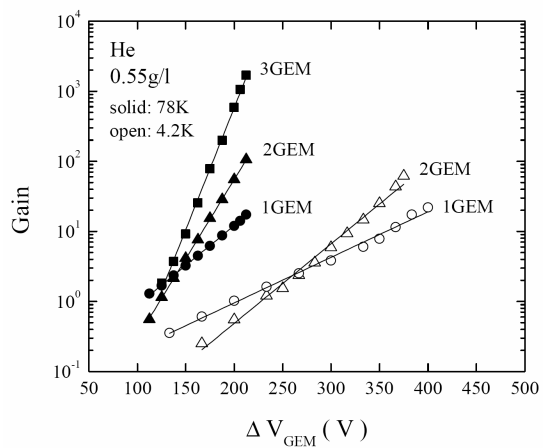

Fig.4 Gain-voltage characteristics of the single-GEM, double-GEM and triple-GEM at 4.2 and 78 K, at a density of 0.55 g/l. At 4.2 K the maximum gains are limited by discharges.

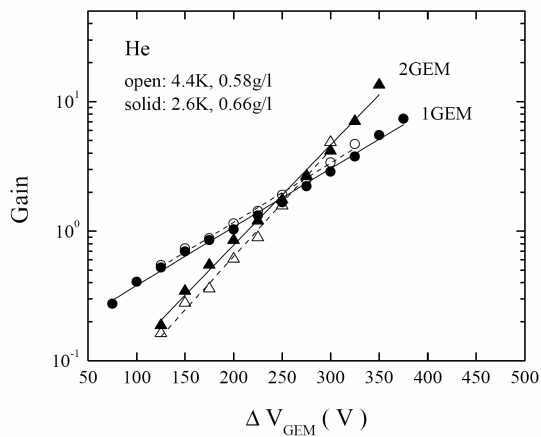

Fig.5 Gain-voltage characteristics of the single-GEM and double-GEM in He at low temperatures, at 2.6 and 4.2 K, at a density of about 0.6 g/l.

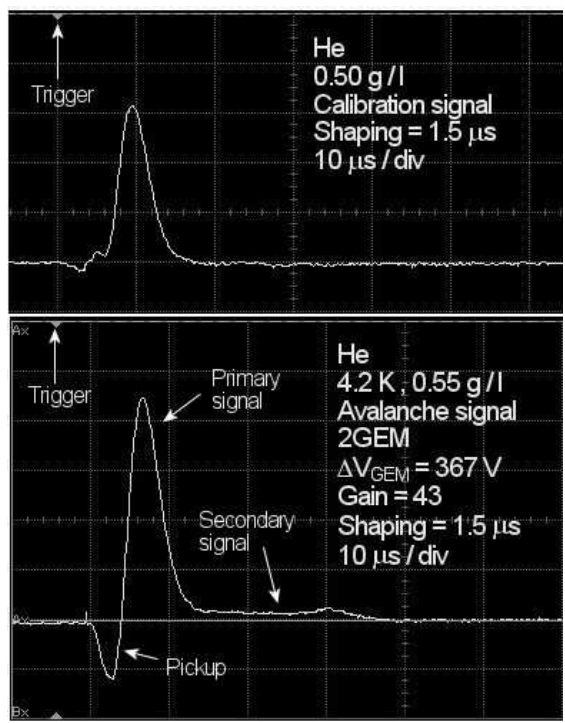

Fig.6 A typical avalanche signal from the double-GEM in He at low temperatures (bottom): at 4.2 K, 0.55 g/l and gain of 43. A calibration signal (from the first GEM acting as an anode) is shown for comparison (top).

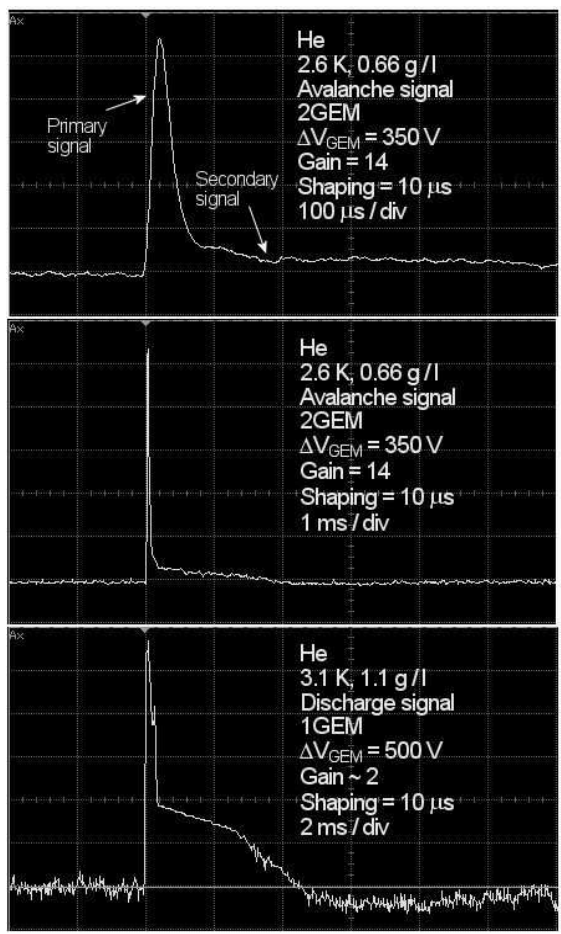

Fig.7 A typical avalanche signal from the double-GEM in He at low temperatures (top and middle): at 2.6 K, 0.66 g/l and gain of 14. A typical "discharge" signal in He at low temperatures is also shown (bottom).

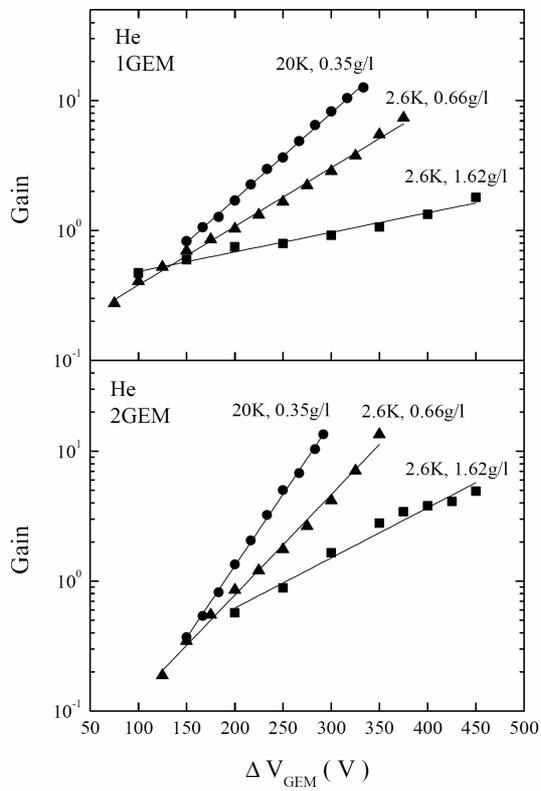
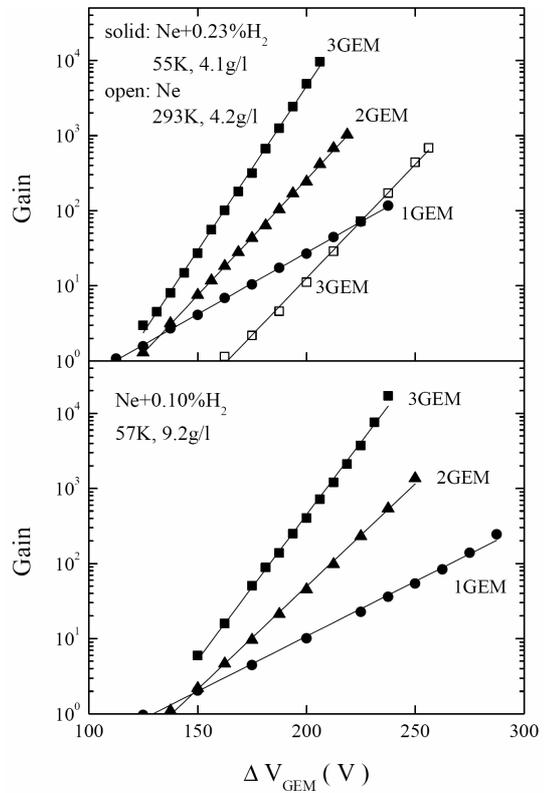

Fig.8 Gain-voltage characteristics of the single-GEM (top) and double-GEM (bottom) in He at low temperatures, at different densities. At 1.62 g/l the maximum gains are limited by discharges.

Fig.9 Gain-voltage characteristics of the single-GEM, double-GEM and triple-GEM in Ne-based mixtures. Top: in Ne+0.2%$H_2$ at 55 K and in Ne at 293 K, at 4.1 g/l. Bottom: in Ne+0.1%$H_2$ at 57 K, at 9.2 g/l.

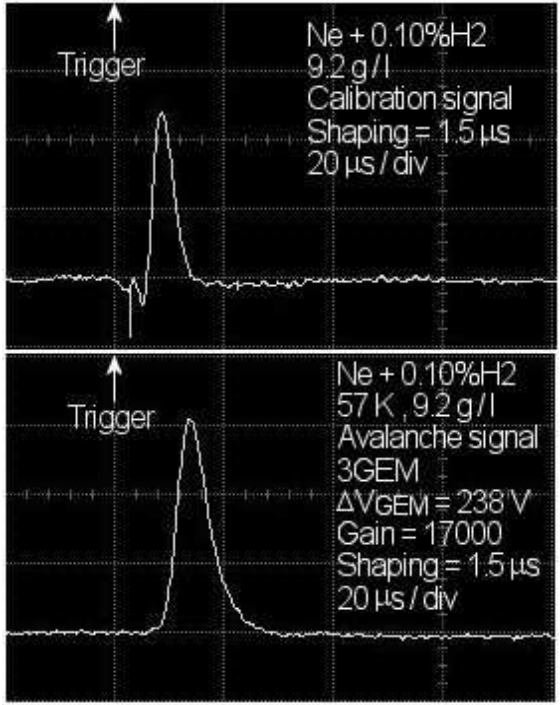

Fig.10 A typical avalanche signal from the triple-GEM in Ne+0.10%H$_2$ at low temperatures (bottom): at 57 K, 9.2 g/l and gain of 17000. A calibration signal (from the first GEM acting as an anode) is shown for comparison (top).

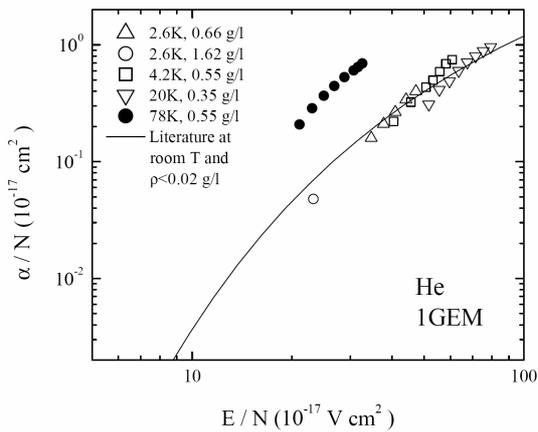

Fig.11 Reduced ionization coefficients as a function of the reduced electric field in He at low temperatures, obtained from single-GEM gain-voltage characteristics. The data are compared to those taken from literature [22], obtained at room temperature and low densities.

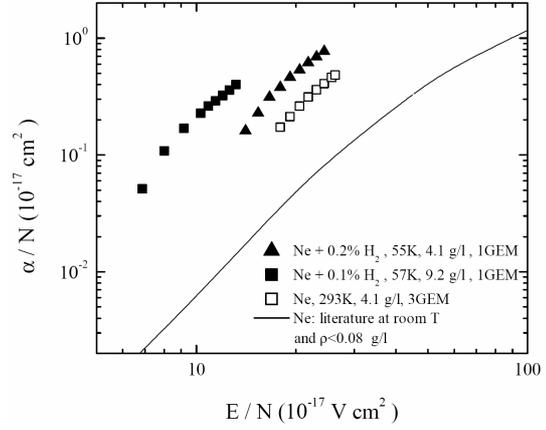

Fig.12 Reduced ionization coefficients as a function of the reduced electric field in Ne+H$_2$, at low temperatures, and in Ne, at room temperature, obtained from single-GEM and triple-GEM gain-voltage characteristics respectively. The data are compared to those taken from literature [23], obtained at room temperature and low densities.

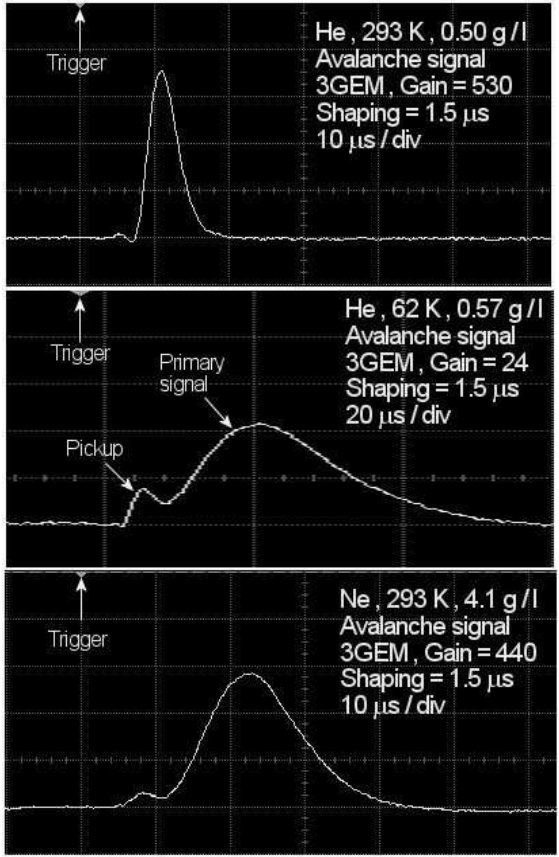

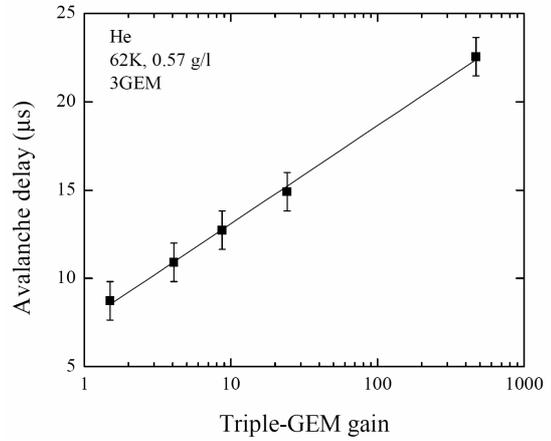

Fig.14 Delay of the avalanche signal with respect to the calibration signal as a function of the triple-GEM gain in He at 62 K. The delay is defined as the peak-to-peak time between the avalanche and calibration signals.

Fig.13 Typical avalanche signals from the triple-GEM in He at room and cryogenic temperatures (top and middle) and in Ne at room temperature (bottom). The figures are adjusted to have the same time-per-cm scale.